\magnification=\magstep1
\line{\hskip 4.2in UCR-HEP-T117\hfill}
\centerline{\bf Time-Symmetric Approach to Gravity}
\vskip .3cm
\centerline{Shu-Yuan Chu}
\centerline{University of California, Riverside, California 92521}
\vskip 1cm

\noindent Abstract: Quantization of the time symmetric system of interacting strings requires 
that gravity, just as electromagnetism in Wheeler-Feynman's time symmetric electrodynamics, 
also be an ``adjunct field" instead of an independent entity.  The ``adjunct gravitational
field" emerges, at a scale large compared to that ot the strings, as a ``statistic" that
summarizes how the string positions in the underlying space-time are ``compactified" into
those in Minkowski space.  We are able to show, {\it without} adding a scalar curvature term
to the string action, that the ``adjunct gravitational field" satisfies Einstein's 
equation with no cosmological term.

\vfill\eject

\centerline{I. Introduction}
In a classic series of papers Wheeler and Feynman showed that their time-symmetric
electrodynamcis, where the advanced interaction is equal in strength to the causal
retarded interaction, is a viable alternative to the usual theory containing only the
retarded interaction [1].  By imposing the complete absorber boundary condition, which
requires that there are enough absorbers to absorb all the radiation in the system, they
demonstrated that the effect of the advanced electromagnetic interaction on a test 
charge is completely canceled, except for the radiation reaction on an accelerated one.
It is therefore a truly surprising result, although perhaps not generally perceived as
such, that Wheeler and Feynman were able to show that the reason Einstein insisted on
pursuing his unified field theory in the face of seemingly insurmountable difficulties
and widespread skepticism perhaps was his firm belief that matter and field should not
be considered as two distinct entities.$^1$  Since matter is the source of field, neither
can exist without the other.  He argued therefore that there is only one reality, which
happens to have two different aspects; and the theory ought to recognize this from the
beginning.  He consistently objected to the highly successful quantum field theories,
where the electrons and the electromagnetic field, for example, are separately quantized
and then coupled together, because he regarded such procedures as improperly doing the
same thing twice.

Between field and matter, his choice in favor of the continuous field as the primary
concept in spite of the discreteness of quantum phenomena no doubt was influenced by
the success of general relativity.  He hoped that the quantum discreteness could be
derived from generally covariant field equations, which ``overdetermine" the fields.$^2$
Although the alternative possibility of eliminating the field as an independent 
entity was actually realized by Wheeler and Feynman in their time-symmetric 
electrodynamcis,$^3$ Einstein commented that it would be very difficult to make a 
corresponding theory for the gravitational interaction.$^4$

Intuitively it seems easier to construct a continuous field from the discrete matter
than the other way around.  At a scale large compared to that of the matter constituents,
the constructed field will appear to be continuous when the underlying discreteness
becomes undetectable.  The purpose of this article is to show that it is possible for
the string theory generalization of the time-symmetric electrodydnamics of Wheeler and
Feynman to construct an ``adjunct" gravitational field this way.  Because the effect
of gravity is so universal and concrete, it may be difficult to imagine gravity not
having its own independent degrees of freedom.  Nevertheless, as we shall see the
elimination of the gravitational field as an independent entity is dictated by the
quantization of the time-symmetric system of interacting strings.

We start with our previous suggestion$^5$ that the quantization of the time-symmetric
system has a statistical orgin.  In a time-symmetric system, such as Wheeler-Feynman
electrodynamcis, the electromagnetic field as already mentioned is not considered
an independent entity with its own degrees of freedom.  It is instead an ``adjunct
field" whose present value is a ``sufficient statistic" that summarizes all the
information about the motion of the charges needed for predicting their future motion.$^6$
As a result, the action of a time-symmetric system of interacting strings is exactly
additive: the string action is a single sum of partial actions, with each partial 
action involving only one string and the value of the ``adjunct (string) field"
evaluated at the position of that string.

The additivity of the string action suggests a connection between the partial action,
which is essentially the area of the string world sheet, and entropy.$^5$  (A possible
connection between area and entropy has been suggested in the different context of
black-hole physics.$^7$)  Based on this connection the classical action has been
identified as the total entropy of the system, the action principle of classical
mechanics as the equilibrium condition that the total entropy of the system be at
a maximum, and the system of interacting strings as being in an equilibrium state of
maximum entropy.  The same connection also suggests that quantum mechanics is the
prescription for calculating the fluctuations of the equilibrium state, with the
path-integral representation of the quantum mechanical density matrix element having 
been shown as an approximation to the partition function of the string theory.$^5$

Because the exact additivity of the classical string action is the key to the
quantization of a time-symmetric system of interacting strings, gravity in such a
system as indicated above must also be an ``adjunct field."  Otherwise, the action
for the free gravitational field will spoil the exact additivity of the string
action.  One therefore expects that the ``adjunct gravitational field" should again 
summarize the information about the motion of the strings, and the curvature of
space-time is just a reflection of the patterns of the tapestry of motion woven
with the world-sheets of the strings.

By studying the free string action, we find that the ``adjunct gravitational field"
in a time-symmetric system emerges, at a scale much larger than that of the strings,
as a ``statistic" that summarizes how the string positions in the underlying space-time
are ``compactified" into those of Minkowski space.  At this larger scale, the classical
equation of motion for the ``adjunct gravitational field" given by the equilibriun
condition of maximum entropy is found to be the Einstein equation.  It should be
emphasized that the Einstein equation is derived here from the kinetic energy term
{\it without} the addition of a scalar curvature term to the string action, in contrast
to the usual practice.  In fact, the addition of any term corresponding to the 
action of a free field is strictly forbidden, because it will destroy the additivity
of the string action.  This restriction leads to the result that there is no
cosmological constant term, which together with the fact that the ``adjunct fields"
have no zero point energy, avoids the difficult problem posed by the stringent
upper bound on the cosmological constant.$^8$
\vskip 1cm
\centerline{II. Derivation of Einstein's Equation}
\vskip .3cm
The classical action for a time-symmetric system of free strings propagating in
an underlying N-dimensional space-time is given by:$^9$
$$S=-T\sum_k\int d^2\sigma_k\vert det(g_{kk})_{ab}\vert^{1/2}\eqno(1)$$
where $(g_{kk})_{ab}=\eta_{mn}\partial_a\xi_k^m\partial_b\xi_k^n$.  In Eq.(1) one
sums over all the strings in the universe in order to ensure that the complete
absorber condition is satisfied.  This condition, which requires that there must
be enough absorbers to absorb all the radiation in the system, maintains macroscopic
causality of the time-symmetric system, whose interaction term necessarily contains
both retarded and advanced interactions.$^3$  The constant $T$ has the dimension of
inverse length squared; $d^2\sigma_k$ is the surface element of the world-sheet 
of string $k$, which is described by specifying $\xi_k^m(\sigma_k^o,\sigma_k^1)$,
the string position at given values of $\sigma_k^0$ and $\sigma_k^1$; $\partial_a
\xi_k^m=\partial\xi_k^m/\partial\sigma_k^a$; the indices $a,b=0,1;\eta_{mn}$ is 
the metric of the underlying space-time; and the indices $m,n=0,...,N-1$.

By the introduction of a metric $(h_{kk})_{ab}$ for the world-sheet of string $k$,
one arrives at an equivalent formula that avoids the awkward square root involving
the $\xi_k$'s:$^9$
$$S=-T\sum_k\int d^2\sigma_kh_{kk}^{1/2}h_{kk}^{ab}\eta_{mn}\partial_a\xi_k^m
\partial_b\xi_k^n,$$
where $h_{kk}^{ab}$ is the inverse of $(h_{kk})_{ab}$ and $h_{kk}$ is the absolute
value of the determinant of $(h_{kk})_{ab}$.  At the extremum of the action, the 
metric$(h_{kk})_{ab}$ is proportional to $(g_{kk})_{ab}$, and the action given
above becomes equal to that given by Eq.(1).  By going to the conformal gauge, 
where $(h_{kk})_{ab}=exp(\phi_k)\eta_{ab}, \eta_{ab}$ is the metric of a flat 
world-sheet and $exp(\phi_k)$ is a conformal factor, we obtain a simplified
string action independent of $exp(\phi_k)$:
$$S=-T\sum_k\int d^2\sigma_k\eta^{ab}\eta_{mn}\partial_a\xi_k^m\partial_b\xi_k^n.\eqno(2)$$

The free string action is a direct generalization of the kinetic energy term of the
particles, with the area of the string world-sheets replacing the length of the
world-lines of the particles.  Because the idea that an ``adjunct gravitational field"
can emerge from such a kinetic energy term is perhaps surprising, it may be worthwhile
to describe in detail the steps involved.

The first step is the usual idea of ``compactification," whereby the positions of the 
strings in the underlying space-time are ``compactified" into the string positions
in Minkow-ski space, $x_k$.  In other words, the $N$ components of $\xi_k$ only
depends on the four components of $x_k: \xi_k^m[x_k^{\mu}(\sigma_k^a)],\mu=0,..,3.$
If Minkowski space were to have the same number of dimensions as the underlying 
space-time instead of the observed four dimensions, then compactification would 
simply be a coordinate transformation in the underlying space-time.

Next, we separate the string position $x_k$ into two parts:
$$x_k(\sigma_k^{\mu})=x_{ko}(\tau_k)+y_k(\sigma_k^a),$$
where $x_{ko}=x_k(\sigma_k^o,\sigma_k^1=0)$ is a particle-like path depending 
on only one variable, which will be denoted by the proper time $\tau_k$; and 
$y_k=x_k(\sigma_k^a)-x_{k0}$.  At a scale much larger than the length of
the string, $y_k$ will appear to be a small string jitter that varies rapidly
about the particle-like path as $\sigma_k^1$ ranges over the entire length
of the string.

The string action given by Eq.(2) can be expressed in terms of the string
positions, $x_k$:
$$S=-T\sum_k\int d^2\sigma_k \eta^{ab}(g_{kk})_{\mu\nu}\partial_ax_k^{\mu}
\partial_bx_k^{\nu},$$
where $(g_{kk})_{\mu\nu}=\eta_{mn}(\partial\xi_k^m/\partial x_k^{\mu})(\partial\xi_k^n/
\partial x_k^{\nu}).$  At a scale much larger than that of the strings, one can
treat the particle-like path of the strings as a continuous variable and
introduce an ``adjunct gravitational field" $g_{\mu\nu}(x_o)$ by interpolating
$(g_{kk})_{\mu\nu}\vert_{y_k=0}.$  The four-dimenisonal space-time metric 
$g_{\mu\nu}$ will have the required ten algebraically independent components,
if the dimension of the underlying space-time is greater than or equal to ten.

Because the number of strings in the universe is necessarily a large number, the
same large number of particle-like paths is included in the sum over the strings.
It should therefore be a good approximation to replace the sum over the strings 
by the functional integral summing all particle-like paths, and we drop the string
label $k$.  The separation of each string position into two parts, as was done 
above, allows us to expand the integrand in powers of the rapidly varying string 
jitters.  Using the requirement of general covariance, we adopt the technique 
used by Polyakov in treating the long-wavelength oscillations of strings and
introduce the covariant derivative in the $y$-direction:$^{10}$ 
$$\nabla_y(\partial_ax_o^{\mu})=y^{\alpha}\nabla_{\alpha}(\partial_ax_o^{\mu})
=y^{\alpha}[\partial(\partial_ax_o^{\mu})/\partial x_o^{\alpha}+\{^{\mu}_{\alpha\rho}
\}(\partial_ax_o^{\rho})],$$
where the Christoffel-bracket
$$\{^{\rho}_{\alpha\beta}\}=(1/2)g^{\rho\sigma}(\partial g_{\sigma\beta}/\partial x_o
^{\alpha}+\partial g_{\alpha\sigma}/\partial x_o^{\beta}  -\partial g_{\alpha\beta}/
\partial x_o^{\sigma}).$$
Because $\{^{\rho}_{\alpha\beta}\}$ is symmetric in the indices $\alpha$ and $\beta$
(absence of torsion), we have
$$\nabla_y(\partial_ax_o^{\mu})=(\nabla_ay)^{\mu}=\partial_ay^{\mu}+\{^{\mu}_{\alpha\beta}\}
(\partial_ax_o^{\alpha})y^{\beta}.$$
The string action expanded to first order of the string jitters is given by:
$$S\approx-T\int D[x_o]\int d^2\sigma\eta^{ab}[g_{\mu\nu}(x_o)\partial_a
x_o^{\mu}\partial_bx_o^{\nu}+2g_{\mu\nu}(x_o)\partial_ax_o^{\mu}(\nabla_by)^{\nu}].$$

The classical equations of motion is given by the action principle, which is the 
equilibrium condition of maximum entropy.  Setting $\delta S=0$, under the infinitesimal
variation $x_o+y\rightarrow x_o+y+\epsilon y'$, we obtain at the lowest order of the
above expansion:$^{10}$
$$\int D[x_o]\int d^2\sigma \eta^{ab}g_{\mu\alpha}\nabla_a(\partial_bx_o^{\mu})
y'^{\alpha}=0,$$
where $\nabla_a(\partial_bx_o^{\mu})=\partial_a\partial_bx_o^{\mu}+\{^{\mu}_{\alpha\beta}\}
\partial_ax_o^{\alpha}\partial_bx_o^{\beta}.$  The equation of motion, expressed in
terms of the proper time, is:
$$d^2x_o^{\mu}/d\tau^2+\{^{\mu}_{\alpha\beta}\}(dx_o^{\alpha}/d\tau)(dx_o^{\beta}/d\tau)=0.
\eqno(3)$$
Eq.(3) describes the particle-like paths of the strings as the geodesics of $g_{\mu\nu}$, 
which justifies the identification of $g_{\mu\nu}$ as the (adjunct) gravitational field.

Using Eq.(3) and the fact that the commutator of two covariant derivatives is the 
curvature tensor, we find at the next order of the above expansion:$^{10}$
$$\int D[x_o]\int d^2\sigma\eta^{ab}[g_{\alpha\beta}(\nabla_ay)^{\alpha}(\nabla_by')^{\beta}
+R_{\alpha\mu\beta\nu}y^{\alpha}y'^{\beta}\partial_ax_o^{\mu}\partial_b x_o^{\nu}]=0.
\eqno(4)$$
The first term in the square bracket gives an equation for the string jittery.  The second
term gives the equation of motion at the large scale that we are looking for.  With the
average (obtained by integrating over $\sigma^1$) of the jitters denoted by: $(1/2)
<y^{\alpha}y'^{\beta}+y^{\beta}y'^{\alpha}>=g^{\alpha\beta}\phi$, where $\phi$ is a
scalar, the equation of motion is:
$$R_{\mu\nu}=0.\eqno(5)$$
Eq.(5) is the Einstein vacuum equation, with no cosmological constant term.  If we had 
included the string interaction term, the righthand side of Eq.(4) would not be zero.
In that case, equating the second term on the lefthand side of Eq.(4) to the terms on
the righthand side which would also be multiplied by the scalar $\phi$, we obtain the
Einstein equation.

In conclusion, we suggest that gravity just as electromagnetism in a time-symmetric 
system is an ``adjunct field," not an independent entity.  It emerges, at a scale much
larger than the length of the strings, as a ``statistic" that summarizes how the string
positions in the underlying space-time are ``compactified" into those in Minkowski
space.  The ``adjunct gravitational field" at the larger scale satisfies Einstein's
equation, with no cosmological term.  The classical string action with gravity
remains exactly additive.  Because of the exact additivity of the string action,
quantization of the time-symmetric system in the presence of gravity vcan proceed 
in the same way proposed previously$^5$ without introducing any new difficulty of
principle.
\vskip .3cm
It is a great pleasure to thank Professors E. Cummings, B.R.Desai, R.H.Dicke,S.Y.
Fung, P.E.Kaus, D.Keane, J.G.Layter, J.G.Nickel, B.C.Shen, G.J.Van Dalen, 
and \break C.H.Woo for many helpful discussions.  This work has been supported by the 
U.S. Department of Energy under grant DE-FG03-86ER40271, DE-AM03-76SF00010, the 
Universitywide Energy Research Group of the University of California, and the 
Energy Science Program of the University of California, Riverside. 
\vskip1cm 
\line{References\hfill}
\noindent 1.  A. Einstein and L. Infeld, {\it The Evolution of Physics}(Simon and
Schuster, 1938), pp.240-243.

\noindent 2.  A. Einstein, {\it Sitzungsberichte}(Preussische Akademie der Wissenschaften, 1923),
p.359.

\noindent 3.  J.A.Wheeler and R.P.Feynman, Rev.Mod.Phys. {\bf 21}, 425 (1949).

\noindent 4.  The comment was quoted in R.P.Feynman, {\it Surely You're Joking, Mr.
Feynman!}(Nor-ton, 1985), p.80.

\noindent 5.  S.Y.Chu, Phys. Rev. Lett., {\bf 71}, 2847 (1993).

\noindent 6.  E.T.Jaynes, in {\it Complexity, Entropy and the Physics of Information,
Santa Fe Institute Studies in the Science of Complexity,}vol.VIII, Ed. W.H.Zurek(Addison-Wesley,
1990),pp.381-403.

\noindent 7.  J.M.Bardeen,B.Carter,and S.W.Hawking, Comm.Math.Phys. {\bf 31},161 (1973);
J.D.\break Bekenstein, Phys. Rev. {\bf D7}, 2333 (1973).

\noindent 8.  S.Weinberg, REv.Mod.Phys. {\bf 61},1 (1989).

\noindent 9.  See, for example, M.B.Green, J.H.Schwarz,and E.Witten, {\it Superstring
Theory}(Cam-bridge University, 1987),p.22.

\noindent 10. A.M.Polyakov, {\it Gauge Fields and Strings} (Harwood Academie, 1987),
pp.253-258.

\end